# Planar defect in approximant: the case of Cu-Al-Sc alloy

Tsutomu Ishimasa[a]*, Yeong-Gi So[b] and Marek Mihalkovič[c]

[a] *Graduate School of Engineering, Hokkaido University, 060 8628 Sapporo, Japan;*

[b] *Graduate School of Engineering Science, Akita University, Akita 010 8502, Japan;*

[c] *Institute of Physics, Slovak Academy of Sciences, 845 11 Blatislava, Slovakia*

**Abstract**

The {110} planar defect formed in the Cu-Al-Sc 1/1 approximant has been studied by means of electron microscopy and 6-dimensional analysis. As the displacement vector of the planar defect, $\frac{1}{2}\langle \tau, \tau^{-1}, 0 \rangle$ has been proposed. Here τ denotes the golden ratio. This vector corresponds to a 6-dimensional translation of type [111000] and also to the *c*-linkage of Tsai-type clusters. An atomic structure model of the planar defect has been constructed referring to the structural properties of the regular 1/1 approximant; namely truncated triacontahedron framework and embedded clusters. This structural model was validated by the agreement between the observed and simulated HAADF-STEM images. Models of the intersection of two planar defects and the triple point, where three planar defects intersect, were also proposed. All local structures formed at these defects are understood by the concept of so-called canonical cell tiling.

*Corresponding Author

E-mail :   ishimasa@eng.hokudai.ac.jp

# 1. Introduction

Since the discovery of quasicrystals [1], their structural features and physical properties have been intensively studied. In the course of these studies, it became important to understand the relationship between quasicrystals and their crystal approximants. The latter is a series of periodic structures consisting of local atomic arrangements similar to the corresponding quasicrystal, and is related to the rational approximation of irrational numbers such as the golden ratio [2]. In particular, defects formed in approximants are thought to provide important hints for understanding structural relationships [3]. For a cubic approximant formed in Al-based alloy, Quiquandon et al. analyzed the {110} planar defect [4]. However, knowledge of such defects is currently very limited.

In this paper, we focus on the planar defect formed in the approximant consisting of Tsai-type cluster. The Tsai-type cluster was first recognized in the Cd-Yb quasicrystal [5]. Structure analysis of this quasicrystal [6] revealed a quasiperiodic framework and embedded atomic clusters. Such structural relationship between the quasicrystal and approximants will be effectively used for constructing atomic structure model of a planar defect.

Tsai-type approximants are formed in many alloys as follows; Cu-, Zn-, Pd-, Ag-, Cd-, and Au-based alloys [7, 8]. In this paper, a 1/1 approximant formed in Cu-Al-Sc alloy [9] is treated as an example. In this system, in addition to the 1/1 approximant, the following quasicrystal-related phases are reported; P-type icosahedral quasicrystal with $a_{6D}$=6.959 Å [10, 11], orthorhombic 1/0-2/1-1/0 approximant with $a$=8.337 Å, $b$=22.02 Å, $c$= 8.305 Å [11, 12].

Before proceeding to the experimental results, we will review the tools needed to analyze defects formed in quasicrystal-related structures.

# 2. Approximant as cluster linkage
## 2.1. Relationship between icosahedral quasicrystal and approximant

The structures of both quasicrystals and their corresponding approximants are understood as cluster linkage. In the case of Tsai-type quasicrystal, triple-shell cluster is embedded in the frame of rhombic triacontahedra, or simply triacontahedra, which are located at the special vertices of Amman tiling named twelve-fold vertex sites [6, 13]. In this quasiperiodic framework, triacontahedra are connected face by face in the two-fold direction, and connected with small overlapping along the three-fold direction (see Ref. 13 for details). The linkages along the two- and three-fold directions are called $b$- and $c$-linkages after Henley

[13]. These linkages are depicted in Figures 1b-d. In the case of *c*-linkage, an oblate rhombohedron occurs at the overlapping part as shown in Figures 1c and d, and thus the triacontahedron is truncated (see, for example, Figure 3e).

The corresponding approximants are interpreted as periodic arrangements of similar truncated triacontahedron but with connection different from the quasicrystal. For example, in the case of 1/1 body-centered cubic (BCC) approximant, the point symmetry is reduced from $m\bar{3}\bar{5}$ to $m\bar{3}$, and the space group is $Im\bar{3}$. Six orthogonal *b*-linkages out of thirty, and eight *c*-linkages out of twenty survive in the 1/1 approximant. Therefore, twenty *c*-linkages fall into two groups, eight allowed and twelve forbidden in the 1/1 approximant. These two types are referred to in the present paper as *c1* and *c2*. They are depicted in Figures 1c and d, respectively and summarized in Table 1. The eight *c1*-linkages correspond to the connection between the body-centered site and the origin of the cubic unit cell, and the six *b*-linkages, named *b1*, to the connection between a corner and the next one. The remaining *b2*- and *c2*-linkages may create a defect in the 1/1 approximant. Such linkage selection errors in quasicrystals are understood by the concept of phason strains [13-15].

## 2.2. Coordinate systems

The coordinate system using six vectors is introduced based on the geometric relationship between the quasicrystal and the approximant [16]. Six vectors $e_i$ ($i$ =1-6) in Figure 1a are the projection of basis vectors in the six-dimensional (6D) hyperspace (see Ref. 17 for details). These vectors are parallel to the lines connecting the vertices and the center of an icosahedron. Note that the length of the vectors is $1/\sqrt{2}$ in the usual projection scheme, but is set to unity here. Using the vectors $e_i$, all the vertex positions of the triacontahedra, including the body-centered one, in the 1/1 approximant are expressed as follows with selected sets of {$n_i$} ;

$$r = a_R \sum_{i=1}^{6} n_i e_i \quad \ldots \quad (1).$$

The parameter $a_R$ is the edge length of the triacontahedron, that is, the edge length of the rhombus on its surface. Similarly, all center positions of triacontahedra are indexed by a set of {$n_i$} by selecting arbitrary center as the origin. In the fourth column in Table 1, 6D description of *b*- and *c*-linkages are presented.

Cubic crystal system [*x*, *y*, *z*] is also introduced that is based on the lattice parameter *a* of the 1/1 approximant. The parameter $a$ is equal to the size of the triacontahedron measured along the $x$-axis, and is related to $a_R$ as follows;

$$a = \frac{2(\tau+1)}{\sqrt{\tau+2}} a_R \quad \ldots \quad (2),$$

where $\tau$ denotes the golden ratio. The parameter $[x, y, z]$ is related to the indices $\{n_i\}$.

$$\begin{cases} x = \frac{1}{2(\tau+1)}\{(n_1 - n_5) + (n_2 + n_3)\tau\} \\ y = \frac{1}{2(\tau+1)}\{(n_4 + n_6) + (n_1 + n_5)\tau\} \quad \ldots (3). \\ z = \frac{1}{2(\tau+1)}\{(n_2 - n_3) + (-n_4 + n_6)\tau\} \end{cases}$$

The six indices, $n_1-n_5$, $n_2+n_3$, $n_4+n_6$, $n_1+n_5$, $n_2-n_3$, $-n_4+n_6$, are equivalent to those proposed in Ref. 18.

The vector of forbidden $c2$-linkage, [111000], shown in Figure 1d is expressed as $\frac{1}{2}[\tau, \tau^{-1}, 0]$ in the cubic system. This vector is divided into two components as follows;

$$\frac{1}{2}[\tau, \tau^{-1}, 0] = -\frac{1}{2}u + \frac{2\tau-1}{2}v \approx -0.5u + 1.12v \quad \text{---} \quad (4).$$

where $u = \frac{1}{2}[\bar{1}\,1\,0]$ and $v = \frac{1}{2}[1\,1\,0]$. This expression may be useful to examine an electron microscopy image observed along the [001] direction, in which projection of the BCC structure has period $a/\sqrt{2}$ in the $[\bar{1}\,10]$ and [110] directions. Note that the vector $\frac{1}{2}[\tau^{-1}, \tau, 0]$ is not a $c$-linkage, because the point group of the 1/1 approximant is not $m\bar{3}m$ but $m\bar{3}$. This vector satisfies the relation;

$$\frac{1}{2}[\tau^{-1}, \tau, 0] = \frac{1}{2}u + \frac{2\tau-1}{2}v \approx 0.5u + 1.12v \quad \text{---} \quad (5).$$

Accordingly, it is impossible to distinguish these two vectors in the [001] projection. This may cause a problem when determining the displacement vector of {110} planar defects.

3. **Experimental procedures**

Starting from the pure elements of Cu (99.99%), Al(99.999%) and Sc(99.9%), an alloy ingot with a nominal composition of $Cu_{45.0}Al_{40.5}Sc_{14.5}$ was prepared using an arc-furnace with an Ar atmosphere. This as-cast ingot includes exclusively the 1/1 approximant. Using a portion of the as-cast ingot, uniformity was checked by an electron-probe microanalyzer (EPMA) JXA-8900M equipped with a wavelength-dispersive type spectrometer at an acceleration voltage 15 kV. This analysis indicated that the composition of the 1/1 approximant is $Cu_{44.2}Al_{40.9}Sc_{14.9}$, which is consistent with the nominal one. The main part of the ingot was sealed in a silica ampoule after evacuation. The ingot was annealed at 802 ºC for 24 h and slowly cooled to room temperature for 12 h. Weight loss during the heat treatment is less than 0.03%.

A powder X-ray diffraction experiment for structure analysis was performed at the beamline BL02B2 in SPring-8 using monochromatic synchrotron radiation with the wavelength $\lambda = 0.70273(4)$ Å. A Debye-Scherrer camera with a radius 286.48 mm was used for the diffraction experiments [19]. The powder specimen was annealed at 353 ºC for 6 h in $3\times10^4$ Pa Ar atmosphere to remove the strain caused by the powdering. Next, the specimen was put in a fine glass capillary with a diameter 0.1 mm. Diffraction intensity was measured at 300 K. The measured intensities were analyzed by the Rietveld method using the software package RIETAN-2000 [20]. The quality of diffraction data was sufficient to analyze the anisotropic displacement parameters at each site. The softwares, Balls & Sticks [21] and VESTA [22], were used for graphical visualization of structure models.

For electron microscopic observation, the specimens were crushed into fragments using an agate mortar and a pestle, and transferred onto a microgrid mesh. Conventional observations, including selected-area electron diffraction, lattice imaging, and dark-field imaging, were performed using a Jeol JEM200CS microscope at an acceleration voltage of 200 kV.

High-angle annular dark-field (HAADF) imaging was performed using a 200 kV scanning transmission electron microscope (Jeol, JEM-ARM200F) with a spherical aberration corrector. The HAADF simulation was carried out using multislice software (HREM Research, xHREM). The following experimental conditions were used in the simulation: The convergence semi-angle of the electron probe was 24 mrad, while the collection semi-angle for the HAADF detector was in the range of 90–370 mrad.

Displacements due to planar defects were analyzed by least squares fitting assisted by fast Fourier filtering. In this analysis, a sin wave-like image, normalized by the magnitude of the wave, was calculated by a single reflection *g*. Appropriate regions were selected in order to avoid the disturbances caused by the periodic processing of the fast Fourier transform. In each region of interest, the contrast of the processed image containing the information of geometric phase θ was fitted by assuming the following function;

$$f(x,y) = A\sin\left\{2\pi \frac{x\cos\alpha - y\sin\alpha}{\lambda} + \theta\right\} + B \quad \text{--- (6)}.$$

Here, the parameters $(x, y)$ indicates the coordinates in the image, and the angle $\alpha$ is adjusted in the exact **g** direction to obtain one-dimensional sin curve from the two-dimensional image. The parameters $A$ and $B$ are ideally unity and zero, respectively. The difference between the wavelength $\lambda$ and the angle $\alpha$ of two adjacent domains separated by a planar defect causes an error in the estimation of the displacement vector.

## 4. Results
### 4.1. Structure analysis by X-ray diffraction

Figure 2a shows the [001] diffraction pattern of the 1/1 approximant. This pattern indicates systematic extinction due to the BCC structure, and also Laue class $m\bar{3}$. The latter is inferred from the intensity difference between the $hk0$ and $kh0$ reflections. Figure 2b presents the powder X-ray diffraction pattern measured by synchrotron radiation. The lattice parameter was determined to be 13.523(3) Å. These results are consistent with the characteristics in the previous report on the 1/1 approximant [9]. Rietveld analysis was performed using the structure model [9] as the starting model belonging to the space group $Im\bar{3}$. Anisotropic displacement was assumed for each site in this analysis, and the atomic compositions and position parameters were modified to better fit the measured intensity distribution. The calculated intensity distribution of the final model is presented in Figure 2b with the deviation from the measured intensity. The reliable factors of the model are $R_{wp}$ = 2.27 %, $R_I$ = 2.66 %, and $R_F$ = 2.17 % showing good fit. On the other hand, when using isotropic displacement, the factor was $R_{wp}$ = 2.89 %. Details of the final model are summarized in Figure 3 and Table 2. The composition of this model is $Cu_{43.5}Al_{42.2}Sc_{14.3}$, which is consistent with both the nominal and the EPMA analyzed compositions. It is noted that the estimated displacement parameters $\{U_{ij}\}$ meet the following requirements for cubic within the range of experimental error;

$$\begin{cases} U_{ii} > 0 \\ U_{11}U_{22} + U_{22}U_{33} + U_{33}U_{11} - U_{12}^2 - U_{13}^2 - U_{23}^2 > 0 \\ \det\{U_{ij}\} > 0 \end{cases} \quad \text{--- (7)},$$

where det{$U_{ij}$} means determinant.

The structure of the 1/1 approximant is interpreted as combination of periodic framework and embedded Tsai-type clusters. The framework consists of the sites M3 and M5. The site M3 is occupied mostly by Cu, and site M5 by Al. This framework has holes at the origin and the body-center position of the unit cell, in which the Tsai-type clusters are embedded. The Tsai-type cluster consists of triple shells plus disordered arrangement (M7) shown in Figure 3a. The triple shells consist of an inner dodecahedron (M2, M4), a Sc icosahedron, and an outermost icosidodecahedron (M1, M6). The atoms consisting of the Tsai-type cluster exhibit interesting fluctuation as presented in Figure 3f. In particular, the fractional site M7 has very large displacement parameters. This is interpreted as a randomly oriented tetrahedron [23]. The characteristic fluctuation observed in Figure 3f may be related to the dynamical motion of atoms reported in the isostructural $Zn_6Sc$ [24, 25] or the static chemical order theoretically predicted in the Cu-Al-Sc approximant [26].

Aside from these detailed structures, the structural interpretation as a framework and embedded clusters will be used below to model planar defects.

## 4.2. Electron microscopy

In the dark-field image shown in Figure 4a, planar defects in the 1/1 approximant are observed. The electron beam is set almost parallel to the [1$\bar{1}$0] direction, and is approximately 0.6 degrees away from the axis. The defect plane was determined to be {110} by analyzing their traces. Intersections and triple points are observed in Figure 4a. At the latter, three {110} planar defects intersect at one line. In the present paper, this will be call "Triple point" instead of "Triple line". An example of the triple point is shown in the rectangle in Figure 4a. An interesting feature of the triple point is frequent occasion comparing with the intersection. In addition, the triple point accompanies no significant image contrast caused by lattice strain. This indicates that the sum of three displacement vectors, $R_i$, around the triple point may satisfy the following condition (see Figure 5b).

$$R_1 + R_2 + R_3 = T \quad \text{--- (8),}$$

where the vector $T$ is the lattice translation of the BCC structure. The sum is not necessarily zero, because a displacement vector is a relative shift between both sides of crystal. In this paper, each displacement vector is set anticlockwise around the triple point. On the other hand, in the case of intersection, the condition of no strain is easily achieved;

$$R_1 + R_2 + R_3 + R_4 = T \quad \text{--- (9).}$$

The vector $R_3$ cancels $R_1$ and $R_4$ does $R_2$, if two planar defects penetrate each other as shown in Figure 5a. Therefore, the intersection of two planar defects is common, but the unstrained triple point is rare in general.

A $[1\bar{1}0]$ lattice image is presented in Figure 6, which corresponds to the region indicated by the rectangle in Figure 4a. In this direction, the (110) planar defect could be observed from the side, and its displacement might be measured. However, in Figure 6 it is hard to recognize the presence of the (110) planar defect that is indicated by the arrow A, and no apparent displacement is detected. This means that the [001] component of the displacement vector $R$ is approximately 0 or 1/2, and the $v$ component is small and close to 0, where $v = \frac{1}{2}[1\ 1\ 0]$.

The [001] component was also confirmed by the extinction condition $R \cdot g =$ integer in dark-field imaging. When reflection $g = 00\bar{6}$ was used for imaging, the (110) planar defects disappeared as shown in Figure 7a. However, they were observed as clear lines in Figures 7b and 7c, where reflections 550 and $\bar{3}\bar{3}4$ were used, respectively. To observe these images, the direction of electron incidence was set to meet systematic conditions such as those in Figure 7d, where each target reflection was excited, and an objective aperture smaller than Figure 4b was used to pass the reflection alone.

The displacement vector due to the {110} planar defect is further examined in the high-resolution HAADF-STEM image presented in Figure 8. Here, the incident electron beam is parallel to the [001] direction. In this image, the defect is observed as a line with a clear displacement in the direction $u = \frac{1}{2}[\bar{1}\ 1\ 0]$. The defect steps from the line A to B in the middle of Figure 8. The displacement vector $R$, projected on the (001) plane, was estimated to be $0.49(3)u + 0.12(1)v$. The $v$-component is very small, and does not contradict to the observation in the lattice image presented in Figure 6. In addition, the estimated vector matches those in the equations (4) and (5). Considering the two cases on the [001] component and the BCC translation, the following two vectors are possible; $\frac{1}{2}[\tau, \tau^{-1}, 0]$ and $\frac{1}{2}[\tau, \tau^{-1}, 1]$. Note that $\frac{1}{2}[\tau^{-1}, \tau, 0]$ is equivalent to $\frac{1}{2}[\tau, \tau^{-1}, 1]$, and $\frac{1}{2}[\tau, \tau^{-1}, 0]$ to $\frac{1}{2}[\tau^{-1}, \tau, 1]$, because the golden ratio satisfies the relation, $\tau = \tau^{-1} + 1$. Among them, the vector $\frac{1}{2}[\tau, \tau^{-1}, 0]$, which is one of the *c2*-linkages, will be adopted as the displacement vector in the following section.

Figure 9 presents another HAADF-STEM image including an intersection of two defects lying on the orthogonal {110} planes. The crystal is separated into four domains as in the model in Figure 5a. The displacement vectors $R_1$ - $R_4$ are listed in Table 3, which were estimated by the following two methods; one is least squares fitting mentioned in the previous section, and the other is "grid line method" in which grid lines were drawn manually as shown in Figure 10a. The region, 247×247 Å$^2$, was treated for the first method, which is slightly larger than Figure 9. For the second method, the local structure near the core was analyzed in the region, 103×103 Å$^2$, which is slightly larger than Figure 10a.

The estimated displacements listed in Table 3 indicate that the planar defects observed in Figure 9 are essentially the same type with the displacement vector $\frac{1}{2}\langle\tau, \tau^{-1}, 0\rangle$. Therefore, it is possible to construct an ideal model of an intersection presented in Figure 10b.

However, the actual structure is slightly complicated than the ideal one. The ($\bar{1}$10) defect goes from the upper right to the lower left, and the position changes along the way in Figure 9; namely step to line D from C, and to A from B after the intersection. Strictly speaking, this is a curved defect. In addition, the displacements estimated by least squares fitting have larger errors comparing with those by grid line method. This may indicate small distortions or directional changes near the intersection. Actually the angle between two vectors $u = \frac{1}{2}[\bar{1}\,1\,0]$ and $v = \frac{1}{2}[1\,1\,0]$ in the domain 2 was 0.2 degrees larger than other three domains. Here, the domain 2 is separated by the displacement $R_1$ and $R_2$ as shown in Figure 5a.

## 5. Discussions
### 5.1. Structure model of the (110) planar defect

This section shows that the vector $\frac{1}{2}\langle\tau, \tau^{-1}, 0\rangle$ is a reasonable choice as the displacement of the {110} planar defect. This is done in two steps. First an atomic model of the defect is constructed, then the validity of the model is checked by simulating a HAADF-STEM image. Recall that each cluster center is related to a lattice point of the 6D hypercubic, and that the cluster arrangement has a one-to-one correspondence to the framework. This modeling also assumes that the defect region retains the structural properties of the 1/1 approximant revealed by the X-ray structure analysis.

Figure 11a presents the arrangement of the clusters near the (110) planar defect with the displacement vector, $\boldsymbol{R} = \frac{1}{2}[\tau, \tau^{-1}, 0]$. In the defect plane, a prolate rhombohedron with the edge length $\frac{\sqrt{3}}{2}a$ appears. The rhombohedron is spanned by two $c1$-linkages and one $c2$-linkage, namely $[0010\bar{1}\bar{1}]$, $[010\bar{1}\bar{1}0]$ and $[111000]$. The last is the displacement vector $\boldsymbol{R}$.

The framework of truncated triacontahedra (Figure 11b) is constructed from the above model. Both $c1$- and $c2$-linkages are related to the truncation of triacontahedra. Therefore, once the linkages are decided, the truncation of each triacontahedron is uniquely determined. The next step is to place atoms corresponding to the sites M3 and M5 in Table 2. The former is at the vertices of the truncated triacontahedron, and the latter at the edge centers. The size of the triacontahedron is assumed to be $a$ = 13.523 Å, and the types of atoms are assumed to be the same as M3 and M5 sites in the 1/1 approximant. One can thus create the framework with atomic decoration shown in Figure 11b.

To complete, Tsai-type clusters were put at the centers of all triacontahedra. Here Tsai-type cluster formed in the 1/1 approximant was used, while the actual shape of the cluster may depend on the details of the truncation.

The remaining problem is the atomic decoration of two small prolate rhombohedra with edge length $a_R$. This type of rhombohedron appears on the defect plane as shown in Figure 11b. Two Sc atoms are placed at two points that internally divide the long diagonal of the rhombohedron into the golden ratio $\tau$. It follows the structure models of the 2/1 cubic approximant in Cd-Ca [27] and the icosahedral quasicrystal formed in Cd-Yb [6]. Note that in the pesent case, Cd corresponds to Cu/Al and Ca or Yb corresponds to Sc. The final structural model thus obtained is presented in Figures 12a-c.

## 5.2. Comparison between observed and simulated HAADF-STEM images

In order to simulate HAADF-STEM images, a hypothetical periodic approximant is constructed, which includes the local structure of the (110) planar defect (structure data of this approximant is available on request). This approximant corresponds to the parallelogram in Figures 8, and the ABCD in Figure 11b. The structure is monoclinic belonging to the space group *P*2/*m* (No. 10) with lattice parameters *a*=38.216 Å, *b*=13.523 Å, *c*=19.124 Å, and β=128.72°. Three lattice vectors are respectively [3330$\bar{2}$0], [0$\bar{1}$110$\bar{1}$] and [0$\bar{1}\bar{1}$121] in the 6D description. The atomic arrangement in the unit cell is presented in Figures 12a-c. The unit cell includes six Tsai-type clusters. It contains 230.9 Cu, 217.1 Al, 76 Sc, a total of 524 atoms. The alloy composition is $Cu_{44.07}Al_{41.43}Sc_{14.50}$, which is very close to the 1/1 approximant. As presented in Table 4, all atoms have the corresponding sites in the 1/1 approximant except for site 9. Part of sites 3 and 5 forms two small prolate rhombohedra that do not appear in the 1/1 approximant. Each prolate rhombohedron includes two additional Sc atoms of site 9 (see Sc pair indicated by A in Figure 12a).

Two kinds of simulated HAADF-STEM images are presented in Figures 12d and e with the observed one in Figure 12f. The latter simulated image includes the site 9 Sc atoms, and the former not. Both simulated images agree with the observed one, while the latter image with additional Sc seems to fit better. This agreement between the observed and simulated images certainly supports the structure model of the (110) planar defect as well as the choice of the displacement vector $\boldsymbol{R} = \frac{1}{2}\left[\tau, \tau^{-1}, 0\right]$.

### 5.3. Ideal structures of the intersection and the triple point

Structure models of the intersection and the triple point are constructed in the same way. Figure 10b presents cluster arrangement at the intersection. Arrows indicate displacement vectors around the intersection, $\boldsymbol{R}_1$ : [111000], $\boldsymbol{R}_2$ : [0$\bar{1}\bar{1}$010], $\boldsymbol{R}_3$ : [$\bar{1}\bar{1}\bar{1}$000], and $\boldsymbol{R}_4$ : [0110$\bar{1}$0]. Special arrangement of clusters occurs at the core of the intersection. That is the rectangle ABCD in Figure 10b, where edges AB and CD are *c2*-linkages, but BC and AD are *b2*-linkages. This rectangle is not perpendicular to the [001] direction. This core structure is actually a column elongated in the [001] direction as depicted in Figure 13c. Figure 10c shows the framework near the intersection. At the center of the intersection, new prolate rhombohedron with edge length $a_R$ appears, which is light blue in Figure 10c. This rhombohedron has the same shape as that caused by the planar defect, but with a different orientation.

Figure 14 presents the cluster arrangement near the triple point. Three planar defects intersect at this triple point. They are represented by the rhombohedra with different orientations. They lie on the (110), ($\bar{1}$01) and (011) planes, and their displacements are $\boldsymbol{R}_1$ :

[111000], ***R***$_2$ : [00$\bar{1}$$\bar{1}$01], and ***R***$_3$ : [$\bar{1}$000$\bar{1}$$\bar{1}$], respectively. The sum of three vectors is equal to [010$\bar{1}$$\bar{1}$0], which corresponds to the BCC lattice vector 1/2[1$\bar{1}$1]. This means that: Firstly, the combination of the three displacement vectors satisfies the strain free condition (8). These displacement vectors then generate the triangular column presented in Figure 13d, which follows $3_1$-screw symmetry. This screw symmetry can be perfectly retained around the triple point, as shown in Figure 14. Therefore, in this model, no lattice strain appears at the triple point. This may be the reason for the frequent formation of triple points. Atomic models of the intersection and the triple point can be constructed using the same procedure as described in 5.1. Detailed research on such defects is an interesting subject for the future.

### 5.4. Canonical cell tiling

The models of planar defects, including the intersection and the triple point, are also supported from the viewpoint of "canonical cell tiling" [13, 28]. Figure 13 summarizes the local arrangements of clusters revealed in the present study. The regular 1/1 approximant is a BCC structure, and only A-type cell appears as shown in Figure 13a. At the {110} planar defect, the prolate rhombohedron with edge length $\frac{\sqrt{3}}{2}a$ appears. This rhombohedron consists of two B-cells and two C-cells (Figure 13b). At the triple point, the column consisting of A-cell appears as shown in Figure 13d. As presented in Figure 10, a D-cell column (Figure 13c) occurs at the intersection of two planar defects. The D-cell is an essential component in icosahedral quasicrystal models [13]. While the A-, B- and C-cell cluster arrangements are found also in the 1/1 or 2/1 approximants, the D-cell was never previously observed in Tsai-type approximants. The relatively rare formation of intersections, compared to the triple point, suggests an energetically disadvantageous situation for the D-cell in the Cu-Al-Sc system. This may be associated with the special situation in the atomic decoration discussed in Ref. 28: That is the contact of four Ca atoms around one Ca atom in the case of Cd-Ca system. Further investigation about the stability of D-cell is certainly needed.

In summary, all local structures formed in the Cu-Al-Sc approximant are explained by the concept of canonical cell tiling.

### 6. Conclusion

The present study shows that the combination of electron microscopy and higher dimensional analysis is powerful for studying the real structure of quasicrystal-related phases, although the approximant itself is a 3-dimensional structure. Using common properties between quasicrystals and approximants, an atomic model of the {110} planar defect has been constructed for the Cu-Al-Sc approximant. This structural model was validated by matching the observed and simulated HAADF-STEM images. Based on this structure model, the local structures of the intersections of planar defects, including the triple point, were also proposed. Finally, the present study demonstrates the importance of seeing complex structures not only from the crystal side but also from the quasicrystal side.

**Acknowledgement**


Part of the experiments was conducted in collaboration with T. Honma and A. Hirao. The aperture for dark-field imaging of Figure 7 was prepared with the help of H. Takakura. The authors appreciate their cooperation. They also thank T. Hiroto for providing useful information. Synchrotron radiation experiments were performed at SPring-8 BL02B2 with the approval of the Japan Synchrotron Radiation Research Institute (No. 2008B1255). Part of this work was supported by Nanotechnology-Platform-Project by the Ministry of Education, Culture, Sports, Science and Technology, Japan, Grant Number JPMXP09A18UT0198. T.I. was supported by Kakenhi Grant-in-Aid No. 19H05818 from the Japan Society for the Promotion of Science (JSPS). M.M was supported by the Slovak Research and Development Agency, Grant No. APVV-19-0369.

**Figure Captions**

Figure 1

Linkages of triacontahedra in Tsai-type quasicrystal and approximant. (a) Six vectors $e_i$ ($i$=1-6) and three dimensional coordinate [$x, y, z$]. (b) *b1*-linkage corresponding to [1110$\bar{1}$0], (c) *c1*-linkage [110001], and (d) *c2*-linkage [111000]. Six vectors $e_i$ projected on to the (001) plane are indicated by each number $i$ in (b). In (b), $e_1$ and $e_5$ vectors are in the (001) plane, but $e_2$ and $e_6$ have positive $z$-components, and $e_3$ and $e_4$ have negative $z$-components.

Figure 2

Diffraction patterns of the Cu-Al-Sc approximant. (a) Selected-area electron diffraction pattern incident along the [001] direction. A: 006 and B: 060 reflections. (b) Powder x-ray diffraction pattern measured at 300 K. Measured and calculated intensities, peak positions, and deviations are displayed in this order.

Figure 3

Structure model of the 1/1 approximant projected along the [001] direction. (a) Partially occupied arrangement in the center. (b) Dodecahedron formed by M2 and M4 sites. (c) Icosahedron formed by Sc atoms. (d) Icosidodecahedron of M1 and M6 sites. (e) Truncated triacontahedron with edge center atoms. (f) Ellipsoids showing anisotropic displacements.

Figure 4

(a) Dark-field image of the 1/1 approximant. (b) Corresponding electron diffraction pattern showing the objective aperture centered at 33$\bar{4}$ reflection. A: 000, B: 00$\bar{6}$, C: 330 reflections.

Figure 5

Model of intersections of planar defects. (a) Intersection where two planar defects penetrate each other. (b) Triple point where three planar defects intersect.

Figure 6

Lattice image of the rectangular region shown in Figure 4a. Three planar defects are

indicated by arrows. The incident electron beam is parallel to $[1\bar{1}0]$ direction. White rectangle drawn at the lower left corner shows the projection of a BCC unit cell, of which size is 13.523 Å × 19.124 Å.

Figure 7

Dark field images of the Cu-Al-Sc approximant. (a) $00\bar{6}$, (b) 550, and (c) $\bar{3}\bar{3}\bar{4}$ images. (d) Diffraction pattern for the image in (a). Each image was observed under such systematic condition using smaller objective aperture than in Figure 4b to pass the target excited reflection alone.

Figure 8

HAADF-STEM image of the 1/1 approximant showing the (110) planar defect. The incident electron beam is parallel to the [001] direction. Square at the upper right corner is a unit cell of the 1/1 approximant.

Figure 9

HAADF-STEM image of the intersection of two orthogonal planar defects. The incident electron beam is parallel to the [001] direction. Arrows indicate positions of the planar defects.

Figure 10

Intersection of two planar defects. (a) HAADF-STEM image of the core region with grid lines. (b) Cluster arrangement in the [001] projection. Two kinds of dots denote difference of $z$-components in the BCC structure; $z = 0$ and $z = 1/2$. Arrows denote displacement vectors $R_1$ -$R_4$. The 6D coordinates are inserted. A: $[1110\bar{1}0]$, B: $[1220\bar{2}0]$, C:$[2320\bar{1}1]$, and D: [221001]. (c) Framework near the core region. A-D correspond to those in (b). Light blue colored prolate rhombohedron occurs, which is oriented differently from the yellow green ones.

Figure 11

Structure models of the (110) planar defect. (a) Cluster linkages at the (110) planar defect. Blue balls denote positions of Tsai-type clusters. A cube drawn by thin lines denotes unit cell

of BCC. Red lines show the prolate rhombohedron spanned by two *c1*- and one *c2*-linkages. (b) Framework of truncated triacontahedra projected along the [001] direction. Notice occurrence of small prolate rhombohedron that looks as two successive rhombi (shaded by yellow green). A: [000000], B: [3330$\bar{2}$0], C: [322101], D: [0$\bar{1}$$\bar{1}$121].

Figure 12

Comparison between atomic models and HAADF-STEM images. (a) - (c) Atomic arrangement in the monoclinic approximant projected along the [0$\bar{1}$0] direction. Directions of lattice vectors *a* and *c* of the cell are indicated by arrows. For color assignment of atoms, see Figure 3. (a) Arrangement of the dodecahedron and the Sc icosahedron. Additional Sc pair is indicated by A. (b) Arrangement of the icosidodecahedron. (c) All atoms in the unit cell. (d) Simulated HAADF-TEM image without the additional Sc pair. (e) Simulated image with the additional Sc pair. (f) Observed HAADF-STEM image corresponding to the parallelogram in Figure 8.

Figure 13

Canonical cells in the perfect and defect structures. Lines connecting cluster centers are *b*- and *c*-linkages. The latter is colored red. (a) BCC structure. (b) Prolate rhombohedron appearing at the {110} planar defect. (c) Column appearing at the core of the intersection. (d) Column appearing at the core of the triple point.

Figure 14

Cluster linkages near the core region of the triple point. Balls denote positions of Tsai-type clusters. Vectors $R_1$-$R_3$ are displacements due to three {110} planar defects. Rhombohedra appearing each planar defect are colored yellow-green. Thin blue lines denote three main axes of the 1/1 approximant. The 6D coordinates of points O - G are as follows; O: [000000], A: [0010$\bar{1}$$\bar{1}$], B: [111000], C: [2220$\bar{1}$0], D: [2131$\bar{1}$$\bar{1}$], E: [322101], F: [100011], G: [110$\bar{1}$01].

Table 1. Cluster linkages in the 1/1 approximant. The *c*-linkages are classified into two types. One satisfies the lattice translation of the BCC, and the other does not as shown in the third column. Multiplicity in the sixth column is number of equivalent linkages due to the point symmetry $m\bar{3}$.

| Type | Fig. 1 | BCC | 6D index | [x, y, z] | Mult. |
|------|--------|-----|----------|-----------|-------|
| *b1* | (b) | Yes | $[1110\bar{1}0]$ | $[1, 0, 0]$ | 6 |
| *b2* | --- | No | $[110\bar{1}01]$ | $\frac{1}{2}[1, \tau^{-1}, \tau]$ | 24 |
| *c1* | (c) | Yes | $[110001]$ | $1/2[1, 1, 1]$ | 8 |
| *c2* | (d) | No | $[111000]$ | $\frac{1}{2}[\tau, \tau^{-1}, 0]$ | 12 |

Table 2. Result of Rietveld refinement of Cu-Al-Sc 1/1 cubic approximant at 300 K. Occupancy of site M7 is 33.3 %. Anisotropic displacement parameters $U_{ij}$ and equivalent isotropic parameter $U_{eq}$ in $10^{-3}$ Å$^2$ are listed in 7-13th columns. For site M7, the requirement det$\{U_{ij}\}>0$ is met, taking into account the experimental error.

| Site | Atom | Set | $x$ | $y$ | $z$ | $U_{11}$ | $U_{22}$ | $U_{33}$ | $U_{12}$ | $U_{13}$ | $U_{23}$ | $U_{eq}$ |
|---|---|---|---|---|---|---|---|---|---|---|---|---|
| M1 | 0.446(6)Cu + 0.554Al | 48h | 0.3411(1) | 0.1997(2) | 0.1117(1) | 12(1) | 16(1) | 29(2) | -4(1) | -9(1) | 12(1) | 19 |
| M2 | 0.517(6)Cu + 0.483Al | 24g | 0 | 0.2455(3) | 0.0934(2) | 10(2) | 68(3) | 15(2) | 0 | 0 | 19(2) | 31 |
| M3 | 0.970(6)Cu + 0.030Al | 24g | 0 | 0.5974(2) | 0.6545(1) | 15(1) | 4(1) | 7(1) | 0 | 0 | 2.8(6) | 9 |
| M4 | 0.38(1)Cu + 0.62Al | 16f | 0.1615(3) | --- | --- | 48(2) | --- | --- | 40(2) | --- | --- | 48 |
| M5 | Al | 12e | 0.1924(5) | 0 | 1/2 | 15(4) | 1(3) | 33(4) | 0 | 0 | 0 | 16 |
| M6 | 0.44(1)Cu + 0.56Al | 12d | 0.4072(4) | 0 | 0 | 16(3) | 63(4) | 10(2) | 0 | 0 | 0 | 30 |
| M7 | 0.193(5)Cu + 0.140Al | 24g | 0 | 0.070(4) | 0.079(2) | 6.3(6)×10$^2$ | 1.2(2)×10$^2$ | 4(1)×10 | 0 | 0 | -7(1)×10 | 2.6×10$^2$ |
| Sc | Sc | 24g | 0 | 0.1872(2) | 0.3049(2) | 4(1) | 4(1) | 9(2) | 0 | 0 | 2(1) | 6 |

Table 3. Displacement vectors due to crossing {110} planar defects. Components along two vectors $u$ and $v$ are listed, where $u = \frac{1}{2}[\bar{1}\,1\,0]$ and $v = \frac{1}{2}[1\,1\,0]$. Components estimated by least squares fitting are shown in the third and fourth columns, and those by grid line method in the fifth and sixth columns.

| Plane | Vector | Least squares fitting | | Grid line method | |
|---|---|---|---|---|---|
| | | $u$-component | $v$-component | $u$-component | $v$-component |
| (110) | $R_1$ | 0.54 (1) | 1.03 (4) | 0.51 (2) | 1.09 (1) |
| ($\bar{1}$10) | $R_2$ | 1.14 (1) | -0.41 (8) | 1.09 (2) | -0.51 (1) |
| (110) | $R_3$ | -0.51 (1) | -1.13 (3) | -0.51 (2) | -1.11 (1) |
| ($\bar{1}$10) | $R_4$ | -1.16 (4) | 0.51 (2) | -1.15 (2) | 0.51 (3) |

Table 4. Hypothetical monoclinic approximant including six Tsai-type clusters. Part of sites 3 and 5 forms a small prolate rhombohedron that does not appear in the 1/1 approximant. There is no Sc atom corresponding to site 9 in the 1/1 approximant.

| Corresponding site in 1/1 ap. | Atom | Number of atoms | Number of atoms in 1/1 ap. |
|---|---|---|---|
| 1 | 0.446Cu+0.554Al | 144 (=48×3) | 48 |
| 2 | 0.517Cu+0.483Al | 72 (=24×3) | 24 |
| 3 | 0.970Cu+0.030Al | 84 (=24×3+12) | 24 |
| 4 | 0.38Cu+0.62Al | 48 (=16×3) | 16 |
| 5 | Al | 40 (=12×3+4) | 12 |
| 6 | 0.44Cu+0.56Al | 36 (=12×3) | 12 |
| 7 | 0.193Cu+0.140Al | 24 (=8×3) | 8 |
| 8 | Sc | 72 (=24×3) | 24 |
| 9 | Sc | 4 | 0 |

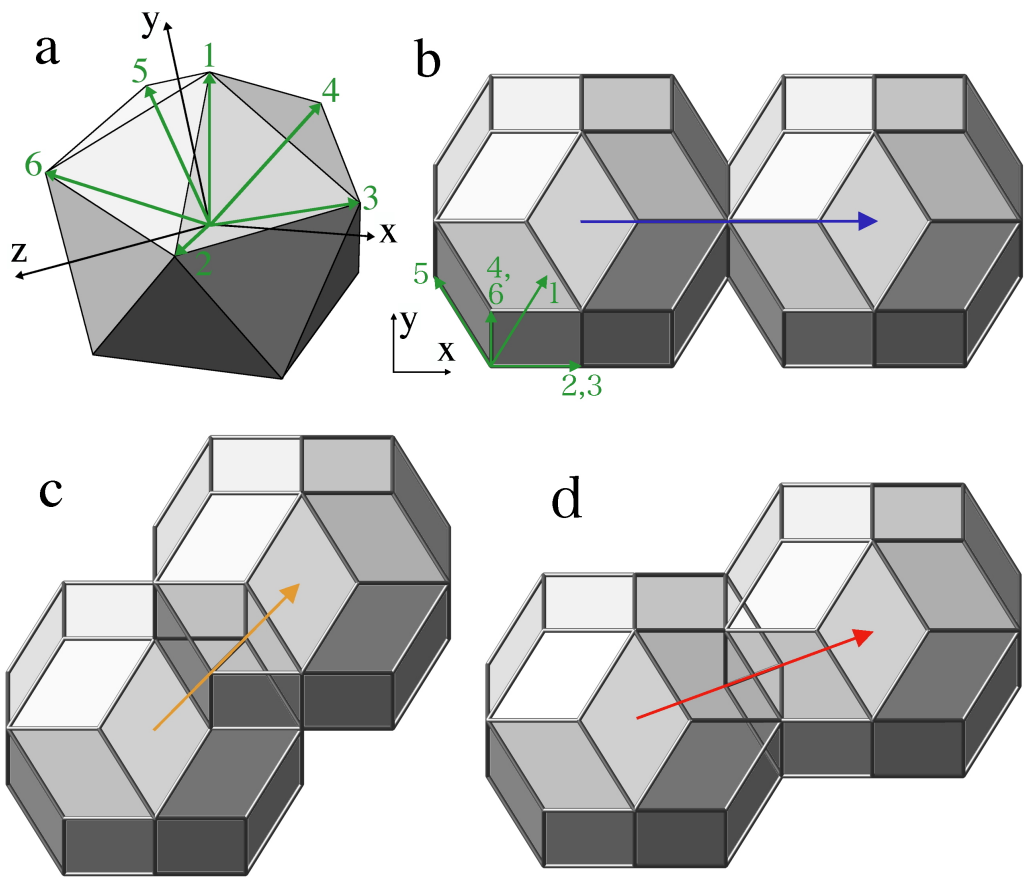

Fig. 1

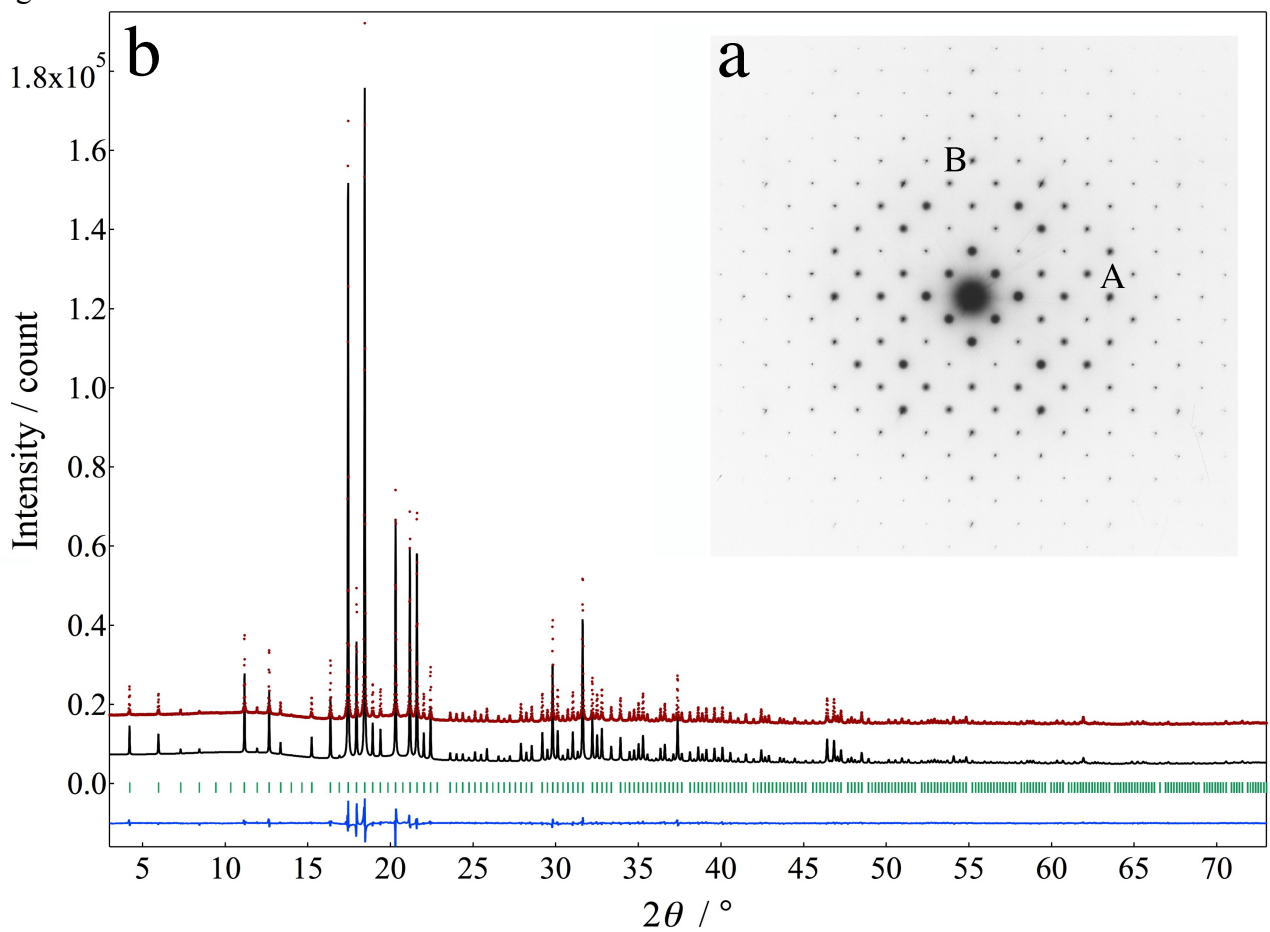

Fig. 2

Fig. 3

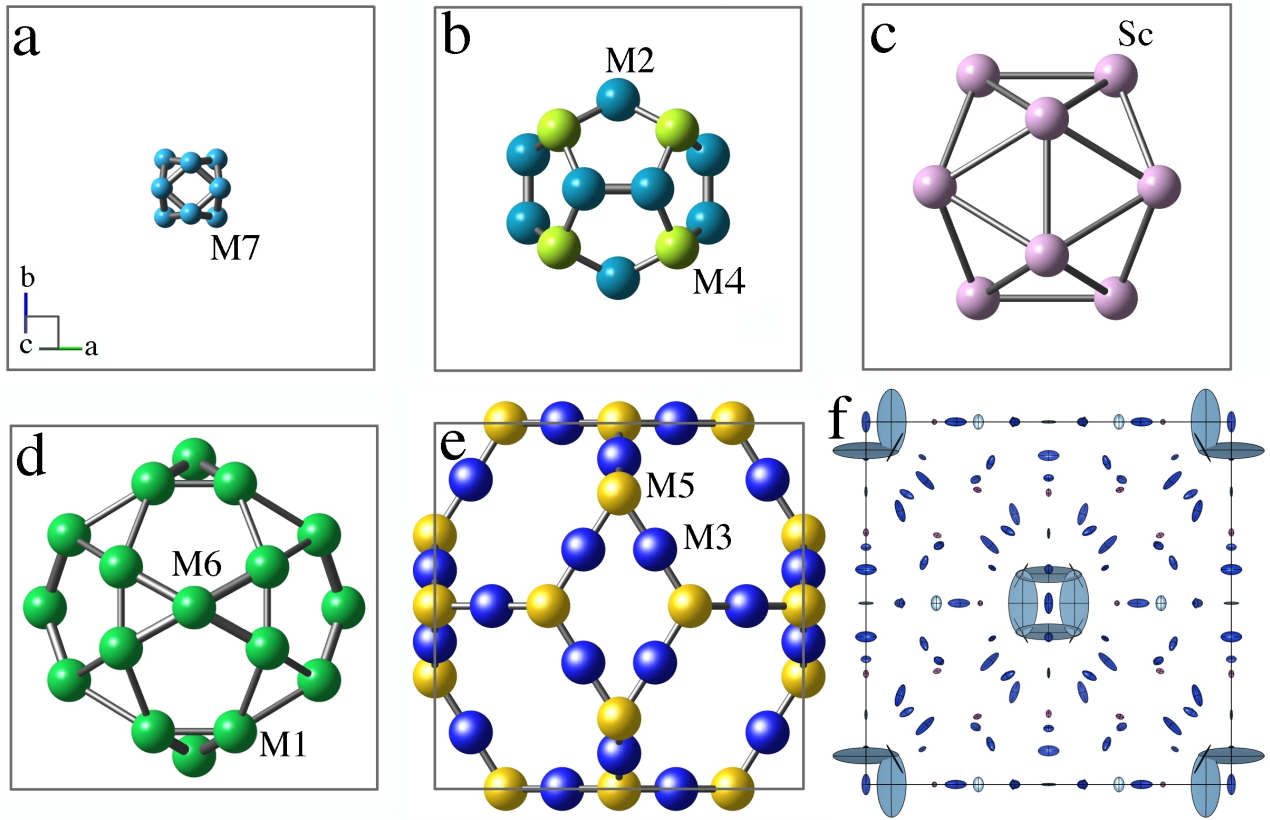

Fig. 4

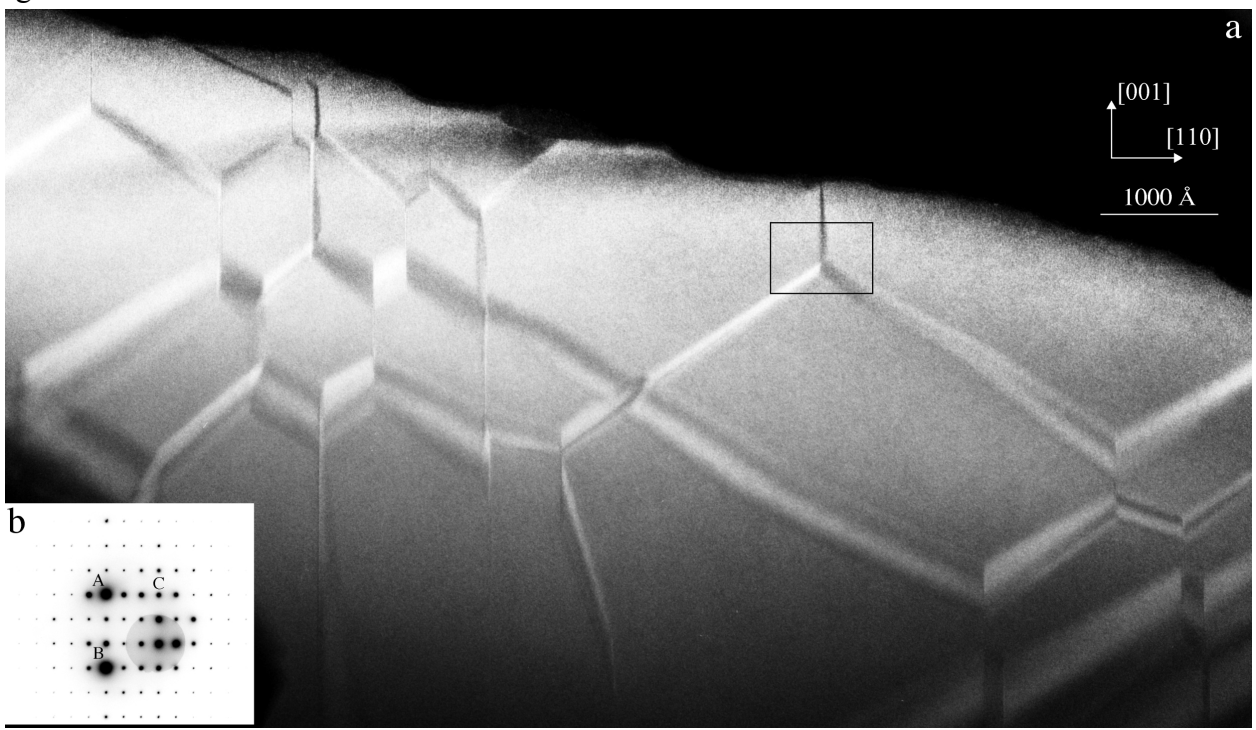

Fig. 5

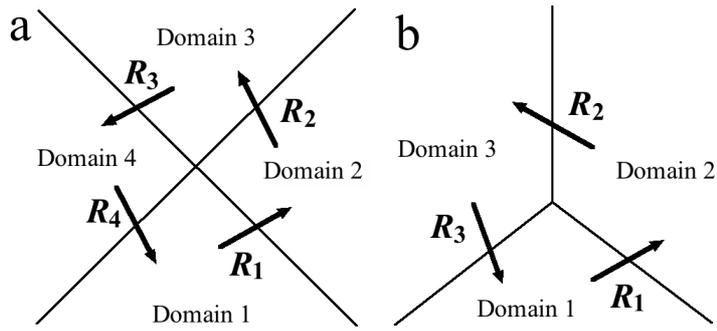

Fig. 6

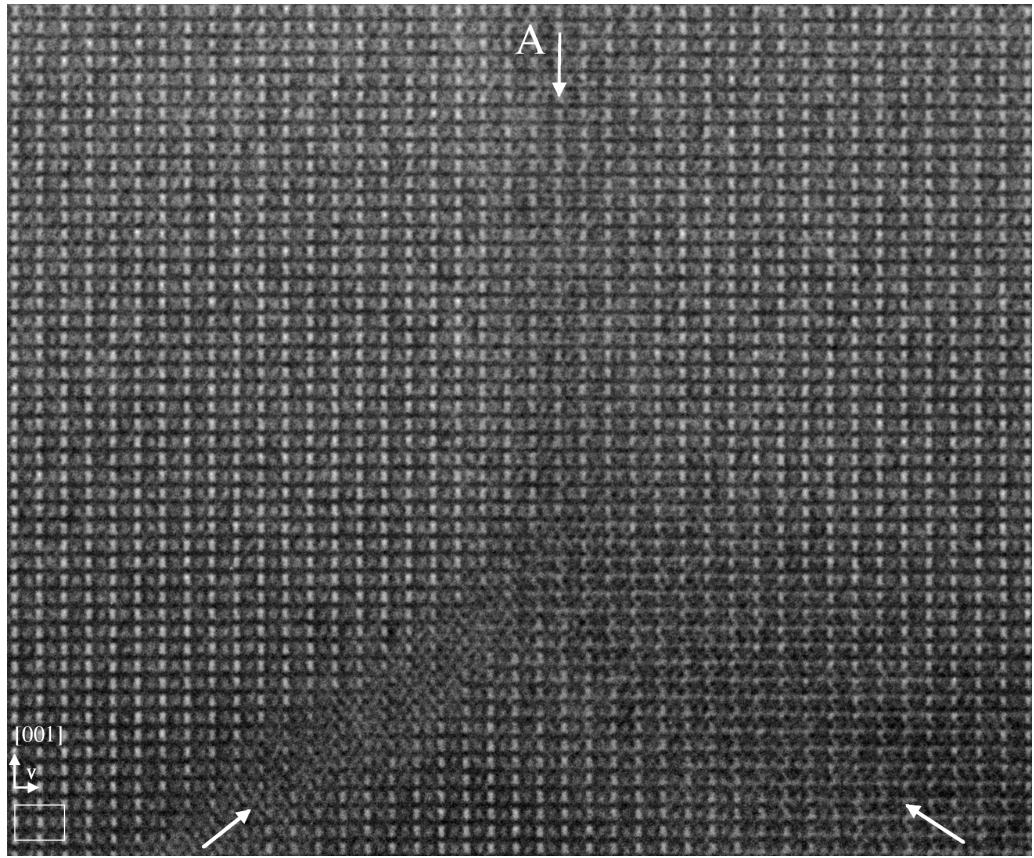

Fig. 7

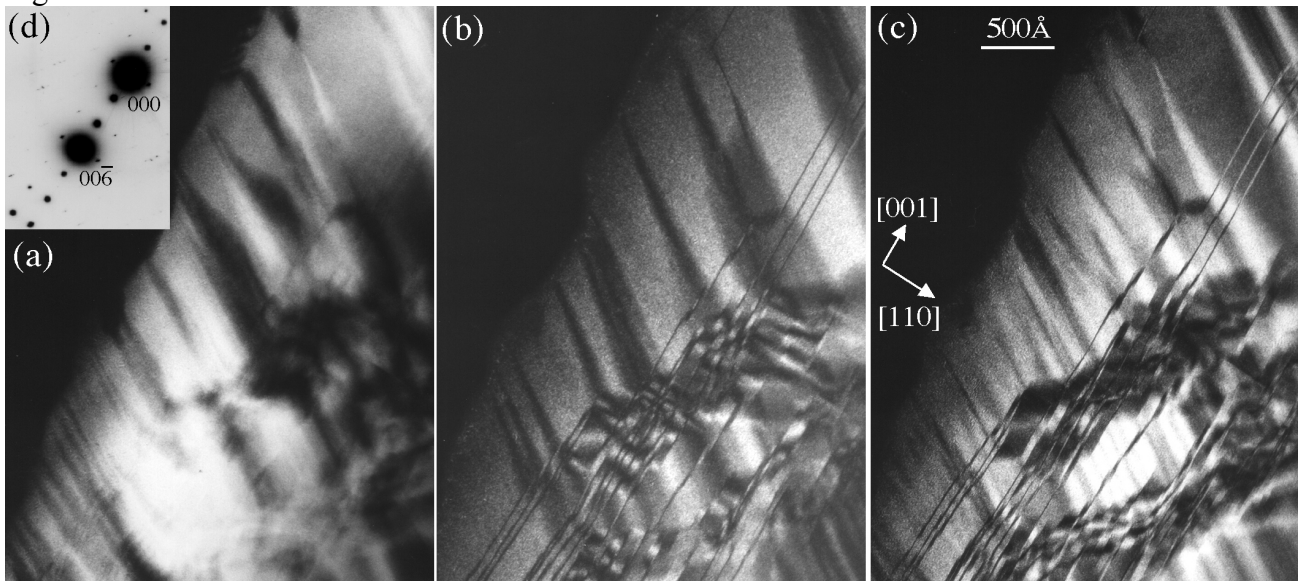

Fig. 8

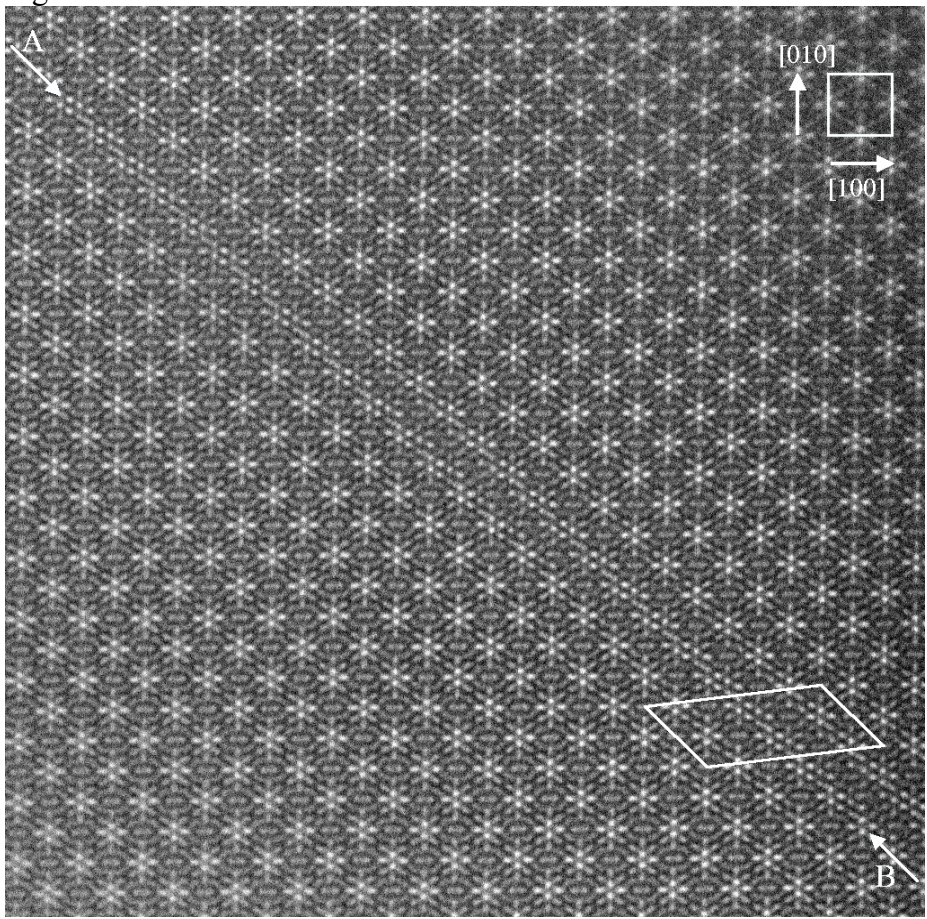

Fig. 9

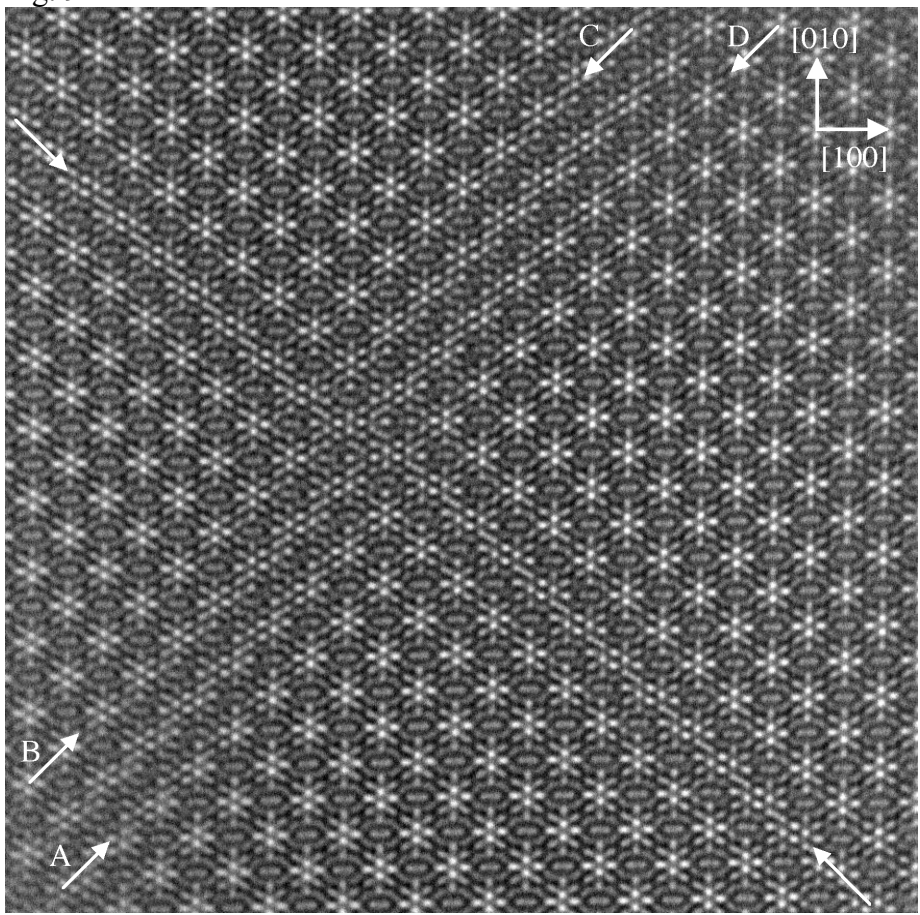

Fig. 10

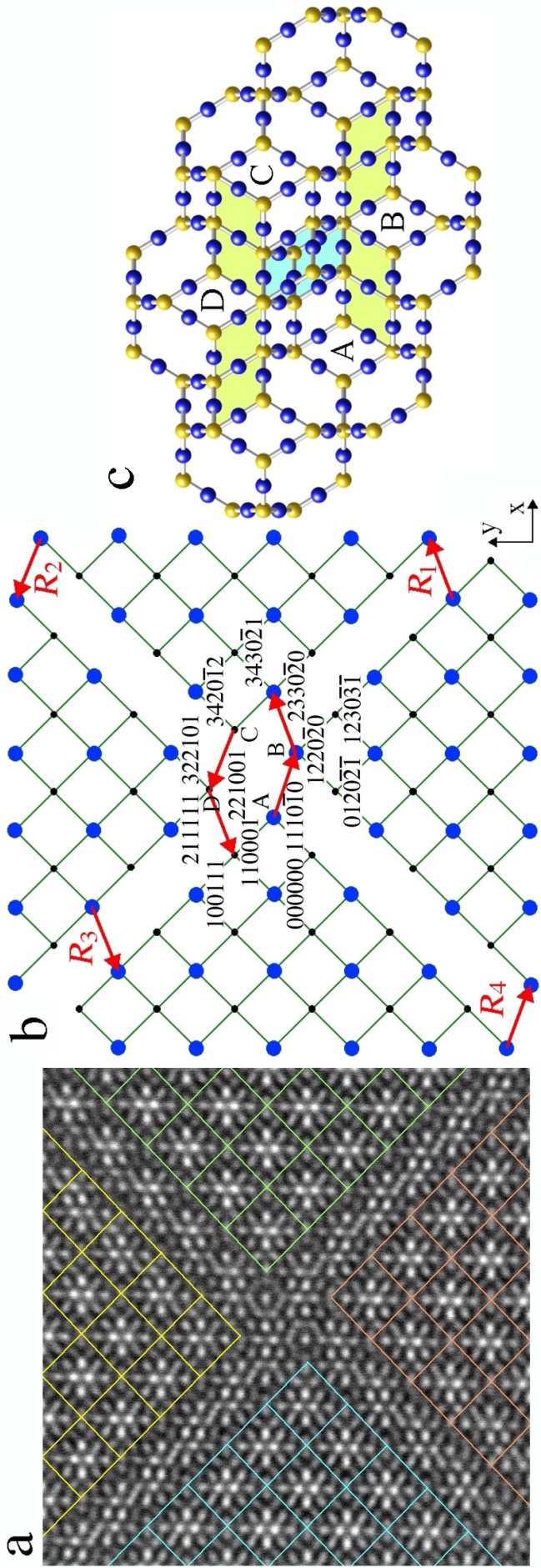

Fig. 11

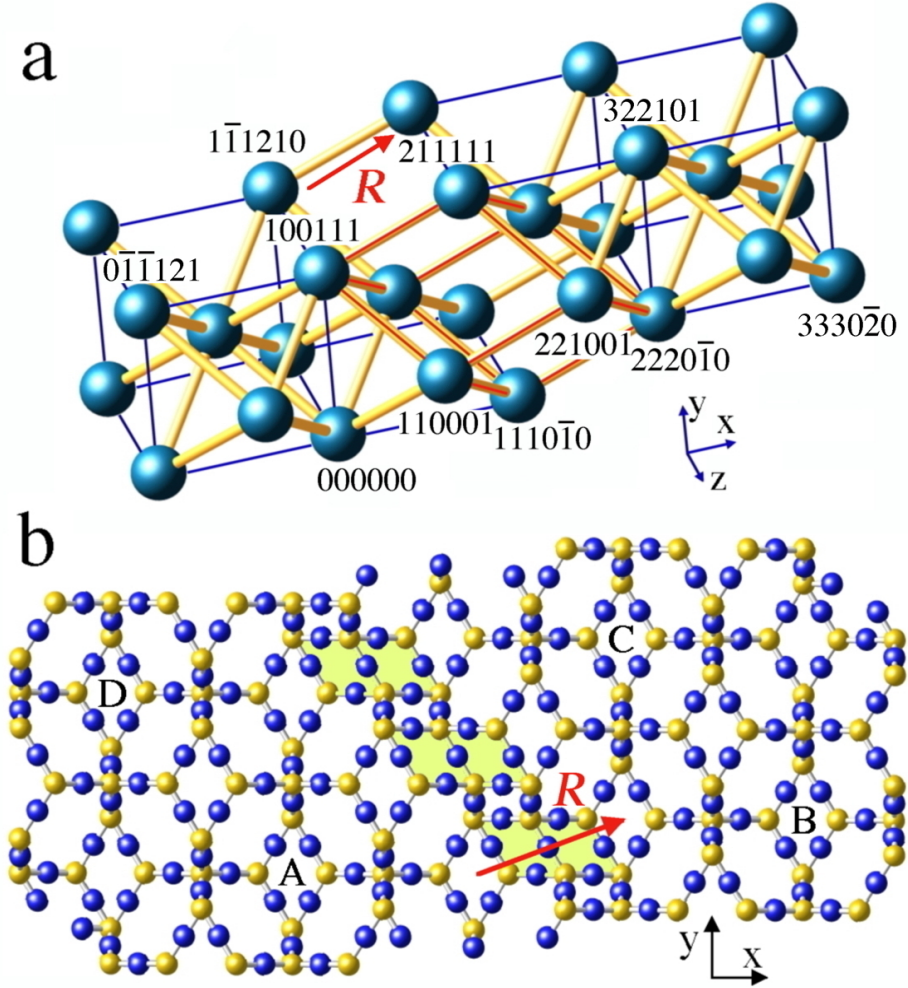

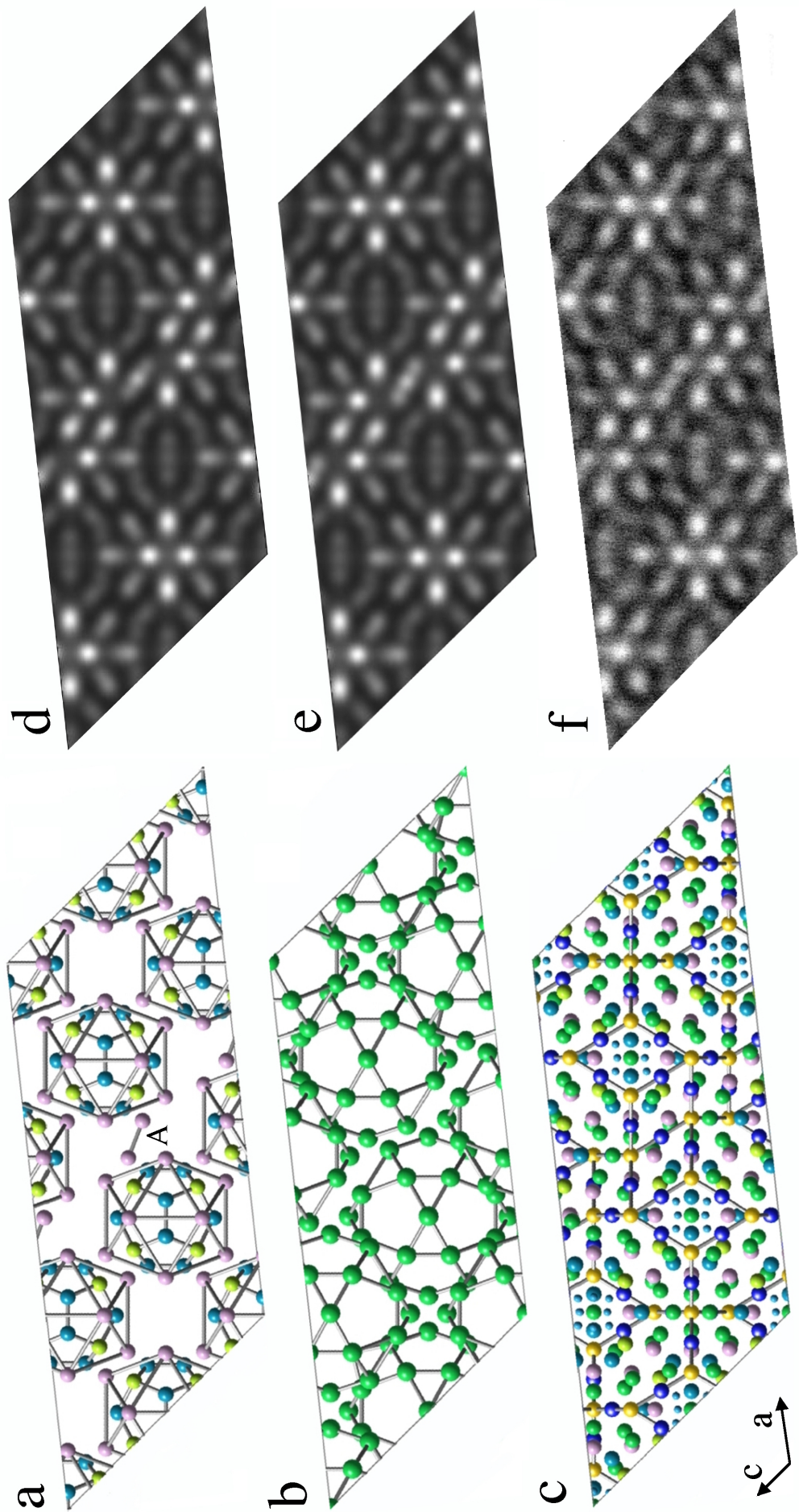

Fig. 13

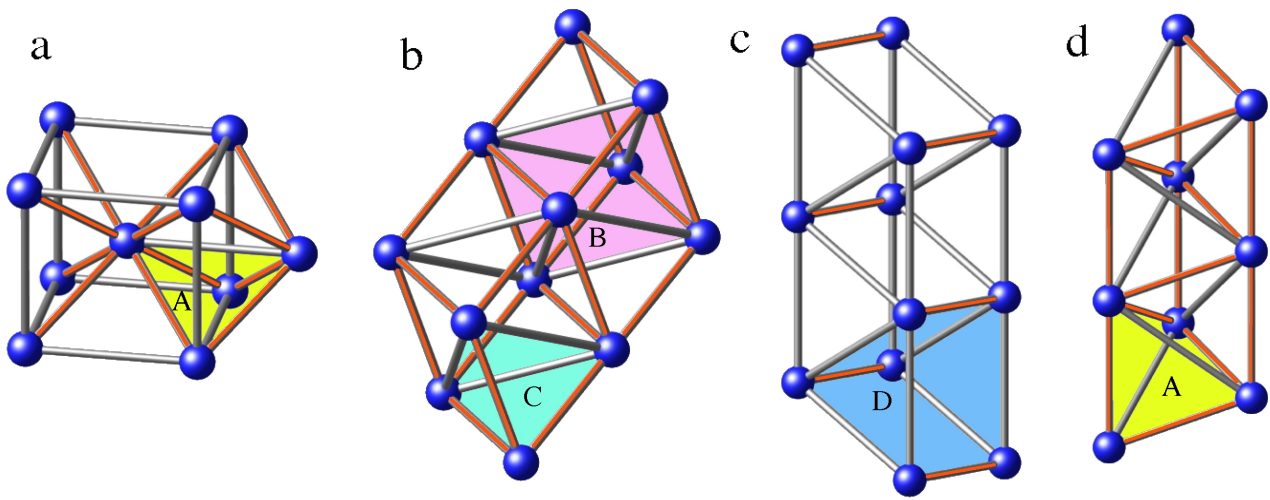

Fig. 14

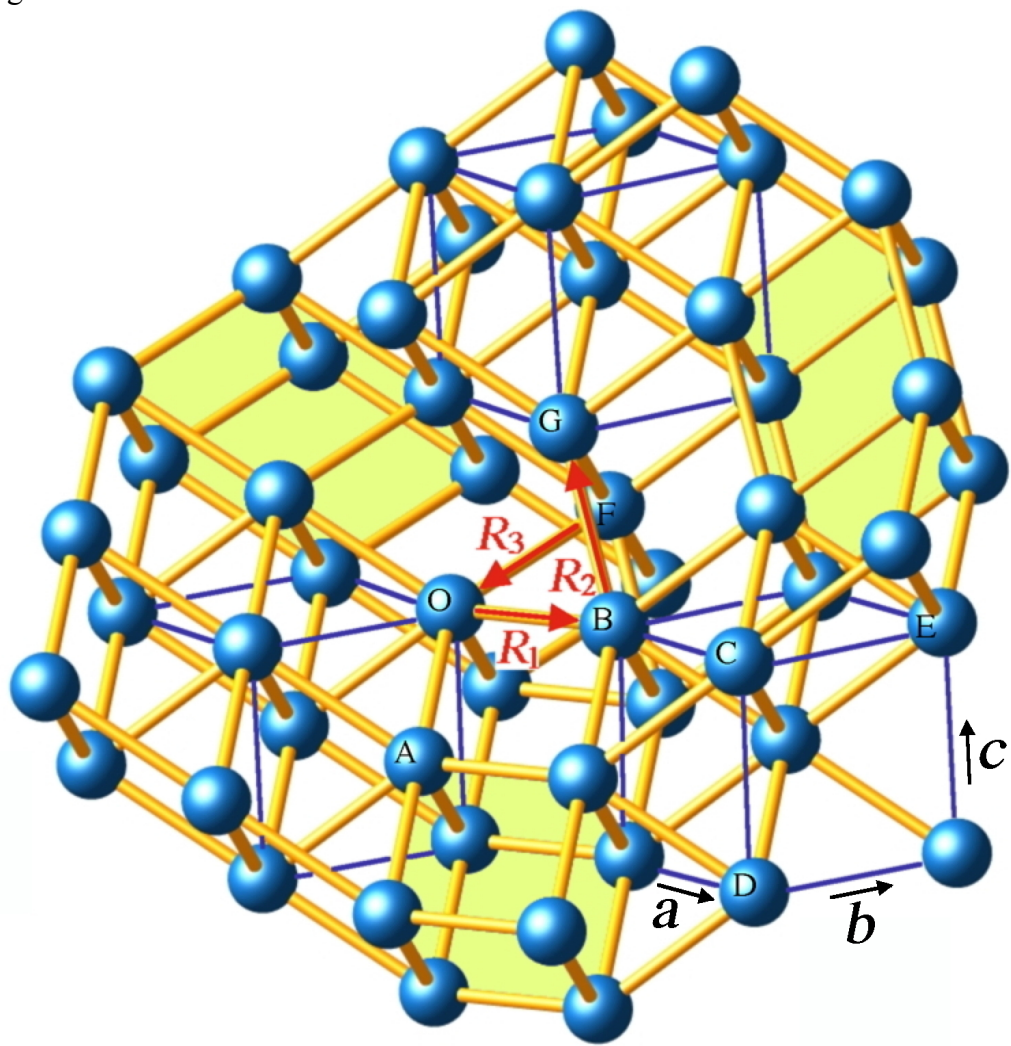